\documentclass[pra,preprint,superscriptaddress]{revtex4}

\usepackage{nicefrac,amsmath,amssymb}
\usepackage{graphicx}
\usepackage{natbib}
\usepackage{bm}
\usepackage{xspace}
\usepackage{stmaryrd}
\usepackage{xifthen}

\newcommand{\mmatrix}[2][]{
  \ifthenelse{\isempty{#1}}
    {\left\llbracket{#2}\right\rrbracket}
    {#1\llbracket{#2}#1\rrbracket}
}

\begin{document}

\title{Real-time nanoparticle characterization through opto-fluidic force induction}

\date{\today}

\author{Marko \v{S}imi{\'c}}
\affiliation{Brave Analytics GmbH, Austria}
\affiliation{Gottfried Schatz Research Center, Division of Biophysics, Medical University of Graz, Neue Stiftingtalstra\ss e 2, 8010 Graz, Austria}
\affiliation{Institute of Physics, University of Graz, Universit\"atsplatz 5, 8010 Graz, Austria}

\author{Doris Auer}
\author{Christian Neuper}
\author{Nikola \v{S}imi{\'c}}
\author{Gerhard Prossliner}
\affiliation{Brave Analytics GmbH, Austria}
\affiliation{Gottfried Schatz Research Center, Division of Biophysics, Medical University of Graz, Neue Stiftingtalstra\ss e 2, 8010 Graz, Austria}

\author{Ruth Prassl}
\affiliation{Gottfried Schatz Research Center, Division of Biophysics, Medical University of Graz, Neue Stiftingtalstra\ss e 2, 8010 Graz, Austria}

% \thanks{Corresponding authors.  E-mail \texttt{christian.hill@medunigraz.at}, \texttt{ulrich.hohenester@uni-graz.at}.}

\author{Christian Hill}
\email{christian.hill@medunigraz.at}
\affiliation{Brave Analytics GmbH, Austria}
\affiliation{Gottfried Schatz Research Center, Division of Biophysics, Medical University of Graz, Neue Stiftingtalstra\ss e 2, 8010 Graz, Austria}

\author{Ulrich Hohenester}
\email{ulrich.hohenester@uni-graz.at}
\affiliation{Institute of Physics, University of Graz, Universit\"atsplatz 5, 8010 Graz, Austria}

\begin{abstract}
We propose and demonstrate a novel scheme for optical nanoparticle characterization, optofluidic force induction (OF2i), which achieves real-time optical counting with single-particle sensitivity, high throughput, and for particle sizes ranging from tens of nanometers to several $\mu$m.  The particles to be analyzed flow through a microfluidic channel alongside a weakly focused laser vortex beam, which accomplishes 2D trapping of the particles in the transverse directions and size-dependent velocity changes due to the optical forces in the longitudinal direction. Upon monitoring the trajectories and velocity changes of each individually tracked particle, we obtain detailed information about the number based particle size distribution.  A parameter-free model based on Maxwell's equations and Mie theory is shown to provide very good agreement with the experimental results for standardized particles of spherical shape.  Our results prove that OF2i can provide a flexible work bench for numerous pharmaceutical and technological applications, as well as for medical diagnostics.
\end{abstract}

\maketitle

\section{Introduction}

Nanoparticles in colloidal suspensions hold significant potential for applications in biotechnology and pharmaceutics~\cite{mitchell:20}.  The use of nanotechnolgy allows for an unprecedented optimization of  nanoparticle-based products, including medical drugs, cosmetics, paper, paints, surface coatings, or lubricants.  To achieve the desired performance, crucial parameters such as size, concentration and, if possible, the shape of nanoparticles must be carefully designed and monitored.  However, the optical characterization of particle sizes and the determination of particle size distributions (PSD) as one of the key particle performance parameters remains a sophisticated and difficult task~\cite{iso:16,anderson:13}.  In dynamic light scattering, the size distribution is obtained from the photon auto-correlation function that decays in time because of the Brownian nanoparticle motion~\cite{finsy:94,stetefeld:16}, whereas in nanoparticle tracking analysis (NTA) the size distribution is obtained by directly monitoring the Brownian motion of individual particles~\cite{hole:13}.  While Brownian-motion based characterization schemes have been established as reliable and flexible offline characterization methods, they hardly provide real-time characterization of nanoparticles, which is drastically gaining relevance in modern production processes.  

Rather than using light for observation purposes only, one can exploit optical forces exerted through momentum transfer between light and matter for the manipulation and analysis of particles, with sizes ranging from several tens of nanometers to micrometers.  This approach has been pioneered by Arthur Ashkin in 1970~\cite{ashkin:70}, and has been awarded the Nobel Prize for Physics 2018.  By using light with an orbital angular momentum (OAM)~\cite{allen:92,franke:08,shen:19} one can additionally transfer angular momentum to matter, which can be exploited for optical vortex trapping of particles~\cite{he:95,gahagan:96,fazal:11}.  Recent research efforts have focused on manifold applications in biology, pharmaceutics, medicine, as well as material science~\cite{grier:03,fazal:11,marago:13,gao:17}, and have established optical tweezers as a most versatile tool in nanoscience, however, with the bottleneck of off-line measurements with a small throughput.

\section{Optofluidic force induction (OF2i)}

In the following we describe and demonstrate a scheme that combines laminar nanoparticle flow through a microfluidic channel with optical-gradient forces, which constrain the location of particle ensembles within a tailored and precisely defined optical field.  For appropriately chosen particle concentrations we can use an ultramicroscopic setup to obtain (i) the intensities from single particle light scattering, (ii) the single particle trajectories, and (iii) the particle number per transported volume.  This information is used to continuously calculate number-based size distributions and concentration, with a throughput of up to 4000 particles per minute, thus overcoming Brownian motion based characterization dynamics used in Nano Particle Tracking analysis and Dynamic Light Scattering.

\begin{figure}[t]
\centerline{\includegraphics[width=0.85\textwidth]{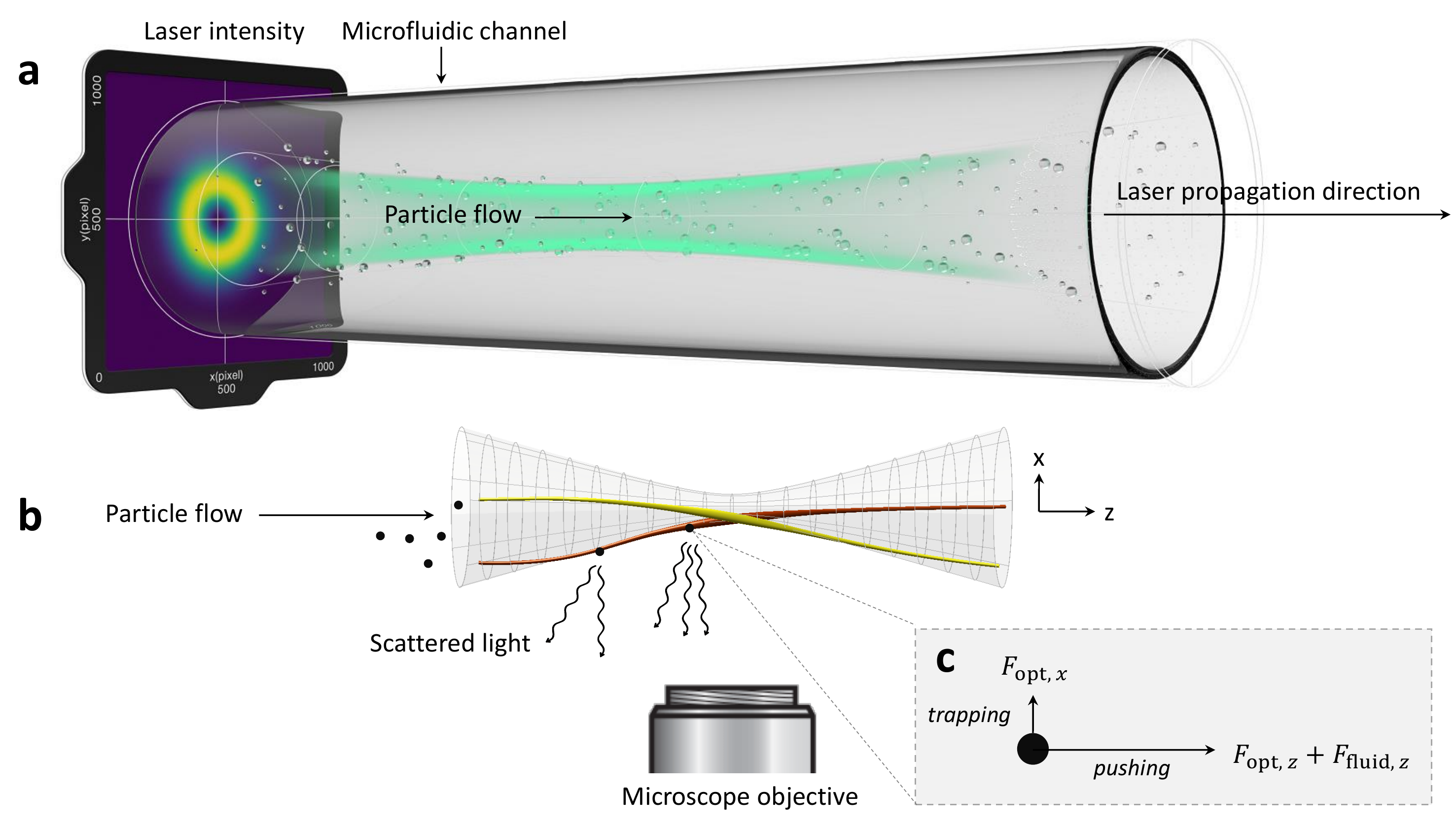}}
\caption{Schematics of the optofluidic force induction (OF2i) scheme.  (a) Particles are immersed in a fluid and are pumped through a microfluidic channel.  A weakly focused Laguerre-Gaussian laser beam with an orbital angular momentum  (OAM) propagates in the same direction as the particle flow, and exerts optical forces on the nanoparticles.  By monitoring the light scattered by the particles through a microscope objective, one obtains information about the scattering cross sections and via particle tracking the velocities of the individual particles.  (b) Simulated trajectories for two selected particles.  Because of the OAM particles move along spiral-shaped trajectories, thus suppressing collisions and particle blockage in the focus region.  (c) The optical force $F_{{\rm opt},z}$ and the fluidic force $F_{{\rm fluid},z}$ acting on a particle control the flow in the propagation direction $z$, the optical force $F_{{\rm opt},x}$ provides optical 2D trapping in the transverse direction $x$ (the trapping force along $y$ is not shown).}
\end{figure}

We denote our approach as opto-fluidic force induction (OF2i)~\cite{hill:20}, for reasons to be described in a moment, whose basic principle is sketched in Figure~1.  The nanoparticles under investigation are immersed in solution and are pumped through a cylindrical flow cell with a parabolic velocity profile, according to Hagen-Poiseuille's law.  In the flow direction (or opposite to it) a weakly focused laser beam propagates along the symmetry axis of the flow cell and exerts optical forces on the nanoparticles.  For appropriately chosen flow and laser parameters we could achieve that particles of different size are brought to a complete halt at different positions of the flow cell, where the fluid and optical forces compensate each other, a technique known as optical chromatography~\cite{imasaka:95,hart:08}.  However, to achieve nanoparticle characterization with a high throughput we seek for a different approach where the particles are consecutively transported through the flow cell.  A critical issue is the focus region of the laser, where the optical forces are largest and particles of different size have different velocities, such that they would block each other or suffer collisions in an uncontrolled fashion.  To overcome this problem, we use an optical vortex beam with an orbital angular momentum (OAM)~\cite{allen:92}, which traps the particles in the transverse directions on the maxima of the doughnout-shaped intensity profile (see Fig.~1) and steers them in the flow direction through the focal region along spiral-shaped trajectories, such that particles of different size can easily pass by each other.  For small particles, the 2D optical vortex trapping ensures a controlled guidance through the flow cell, where an ultramicroscope setup is used to observe the scattered light of the particles.  For larger particles, we can use the same setup to monitor the particle trajectories and infer from the velocity changes due to optical forces the nanoparticle sizes.  With this scheme we combine high throughput with light scattering and trajectory mapping for each nanoparticle, which provides a detailed information for a real-time particle characterization method.

% \subsection*{Experimental setup}

The experimental setup for OF2i measurements consists of a 2D optical trap in a cylindrical flow cell using a weakly focused doughnut shaped vortex beam.  The laser beam is generated by a 532 nm linear polarized CW DPSS laser (Laser Quantum, GEM532) with a maximum output power of 2 W.  The beam alignment is achieved using two mirrors and a 5x expander, see also Appendix A.  A Laguerre-Gaussian mode with topological charge $m=2$ is generated using a vortex phase plate. The ultramicroscope consists of a 10x PLAN microscope objective, an optical filtering bank and a 50 mm focusing lens, and a CCD camera for imaging.  The microfluidic flow cell consists of a continuous, dead volume optimized pumping and laminar fluid handling setup, following derivable fluid continuity principles.  A computer controls laser output power, fluidic flow in the flow cell and records images from a CCD camera over time.  Raw-data evaluation is performed with BRAVE Analytics proprietary software suite H.A.N.S. 2.3 and correspondingly developed Matlab routines.

\begin{figure}[t]
\includegraphics[width=\textwidth]{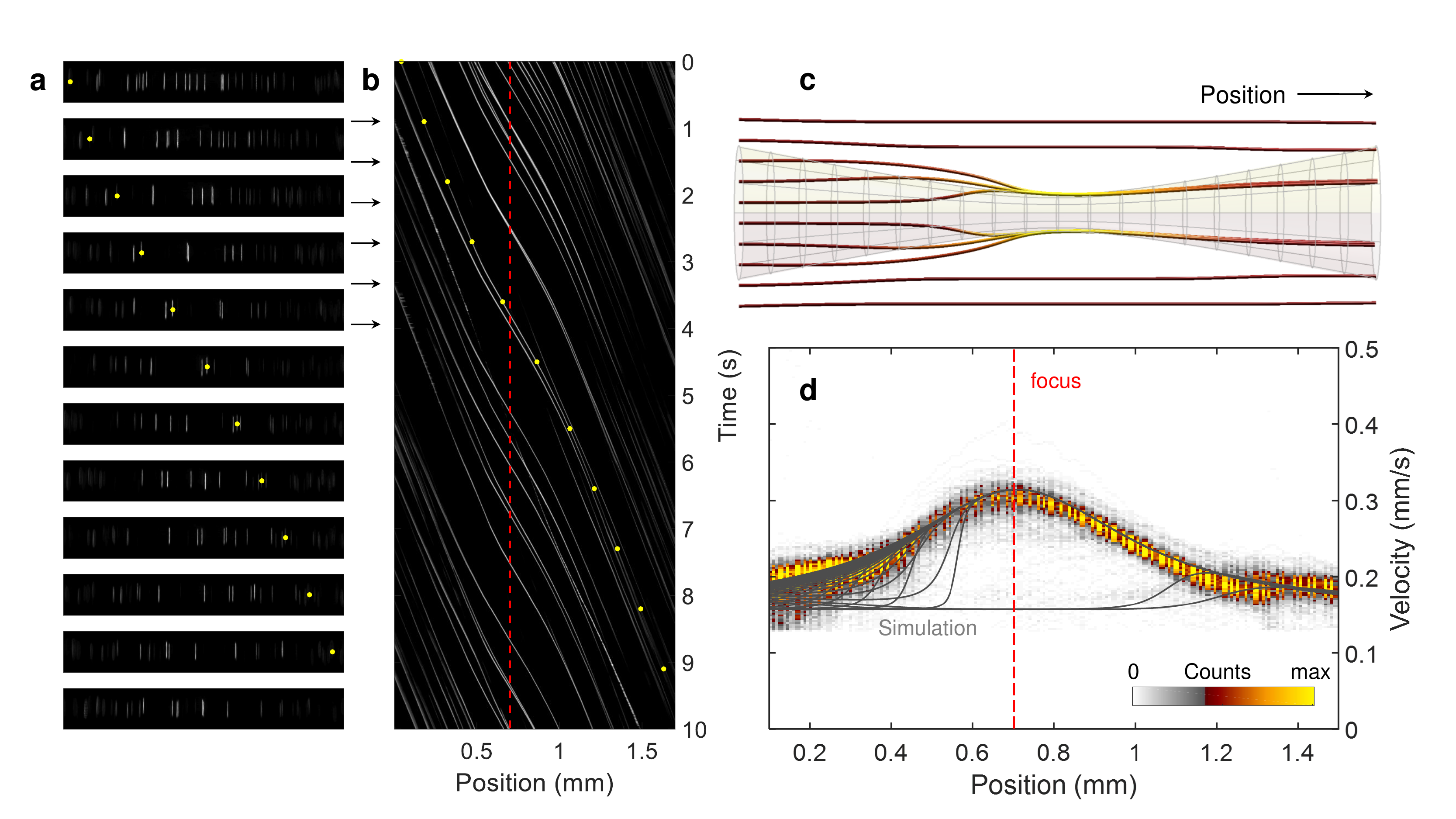}
\caption{Results of OF2i measurements for standard poly latex nanospheres with a diameter of 400 nm.  (a) Selected images of scattered light of the nanoparticles in the flow cell as imaged through our ultramicroscope setup.  By integrating over the transverse (vertical) direction we obtain (b) a waterfall diagram that clearly shows the trajectories of the individual particles, see dots for a selected trajectory.  From the slope of the trajectories we obtain the nanoparticle velocities, which are largest in the focus region (dotted line) because of the optical forces exerted by the vortex beam.  (c) Simulated trajectories using a theoretical model for Mie scattering of Laguerre-Gaussian vortex beams.  (d) Histogram of measured velocities as a function of propagation distance.  The solid lines report the simulation results.}
\end{figure}

\begin{figure}[t]
\includegraphics[width=0.7\textwidth]{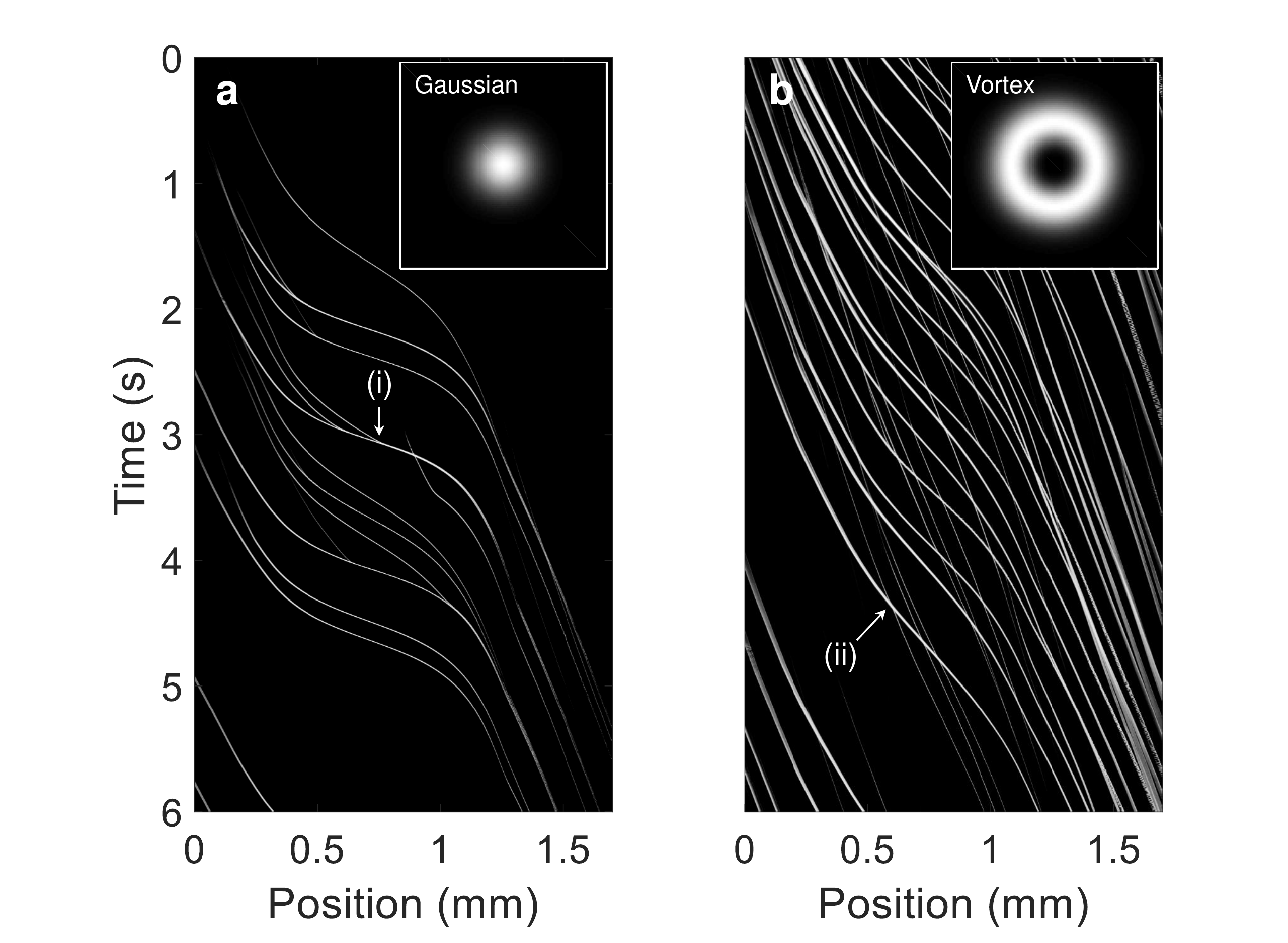}
\caption{Trajectories of individual particles for a mixture of latex nanospheres with diameters of 400 nm, 600 nm, and 900 nm, and for focused (a) Gaussian and (b) Laguerre-Gaussian laser beams, using the laser parameters given in the text with topological charges of zero and two.  For the Gaussian beam one observes frequent collisions between individual particles, see for instance event marked with (i), whereas for the Laguerre-Gaussian beam collisions are strongly suppressed because particles can pass by each other in the focus region of the laser, see event (ii) for an avoided crossing.}
\end{figure}

% \subsection*{Trajectory mapping of polystyrene particles}

We demonstrate the OF2i working principle using standard polystyrene particles with  well-known size distributions with very narrow standard deviation, for details see Appendix A.  Figure 2a shows for nanospheres with a diameter of 400 nm selected images of the CCD camera, which records with a rate of 200 fps the light scattered off the particles.  In our setup, the scattered light becomes diffracted by the glass wall of the flow cell, which acts as a cylinder lens and images the nanoparticles (located in the focus region of the lens) approximately as lines~\cite{meinert:17}.  We next integrate in each image the intensity along the direction perpendicular to the flow axis in order to obtain the waterfall plot shown in Fig.~2b, which clearly exhibits the trajectories of the individual nanoparticles.  The dots in panels (a) and (b) indicate the trajectory of a selected particle.  Figure 3 compares trajectories of an ensemble of nanoparticles with different diameters and for a focused (a) Gaussian and (b) Laguerre-Gaussian laser beam with topological charges of zero and two, respectively.  For the Gaussian laser beam we observe frequent collisions between particles of different size, see for instance event (i), whereas such collisions are strongly suppressed for the Laguerre-Gaussian beam, see event (ii) for an avoided crossing.

From the slope of the trajectories we can compute the nanoparticle velocities at each position of the flow cell, for details see Appendix A.  Figure~2d shows for each position a histogram of the velocities recorded over a time interval of about two minutes.  Away from the focus region the particles move with the velocity of the fluid ($\approx 0.2$ mm/s), and become accelerated and decelerated in the focal region by the optical forces of the vortex beam.  As more and more particles become trapped by the beam while being transported down the fluidic channel, the number of observed velocity counts increases with increasing propagation distance.

\section{Theory}

To acquire a better understanding of the observed results, we have developed a theoretical model for Mie scattering of a spherical nanoparticle excited by a Laguerre-Gaussian laser beam.  Within Mie theory~\cite{bohren:83}, one performs a multipole expansion of the electromagnetic fields in terms of spherical Bessel and Hankel functions, as well as vector spherical harmonics~\cite{bohren:83,jackson:99,hohenester:20}.  The electromagnetic fields outside of the nanosphere can then be represented by the multipole coefficients $a_n$, $b_n$, where $n$ labels the different spherical degrees and orders.  Our theoretical approach for computing the optical forces can be broken up into three steps.  First, we compute the coefficients $a_n^{\rm inc}$, $b_n^{\rm inc}$ for the incoming fields of the Laguerre-Gaussian beam following the approach of~\cite{kiselev:14}.  In a second step, we compute the coefficients $a_n$, $b_n$ for the scattered electromagnetic fields by employing the usual Mie coefficients~\cite{bohren:83,jackson:99,hohenester:20}.  Finally, from the knowledge of the total electromagnetic fields, this is the sum of incoming and scattered fields, we compute the force acting on the spherical nanoparticle using the analytic expressions given in~\cite{gutierrez-cuevas:18}.  As input for our theoretical model we require the nanosphere radius $R$, the refractive indices of the nanosphere and the embedding medium, as well as a parameterization of the electromagntetic fields of the incoming laser beam.  The values used in this work are listed in Appendix A.  From our model we then obtain as output the optical force $\bm F_{\rm opt}(\bm r)$ acting on the nanosphere at some position $\bm r$.

In a microfluidic channel, the velocity $\bm v=\dot{\bm r}$ of the nanosphere with radius $R$ and mass~$m$, which might include the added mass of the fluid, can then be obtained from Newton's equations of motion $m\dot{\bm v}=\bm F_{\rm opt}(\bm r)-6\pi\mu R(\bm v-\bm v_{\rm fluid})$, where the second term on the right-hand side is Stokes' drag accounting for the force of viscosity on the sphere moving through a viscous fluid, which flows with velocity $\bm v_{\rm fluid}$ and has the dynamic viscosity $\mu$.  For sufficiently large spheres, say for diameters above 10 nm, the momentum relaxation time is so short that we can approximately set $\dot{\bm v}\approx 0$~\cite{neuman:08,bui:17}.  The nanosphere's velocity $\bm v$ is then obtained from the condition that the optical force is balanced by the drag force, and we get 
%for a laser propagating in the same direction as the fluid

\begin{figure}[t]
\centerline{\includegraphics[width=\textwidth]{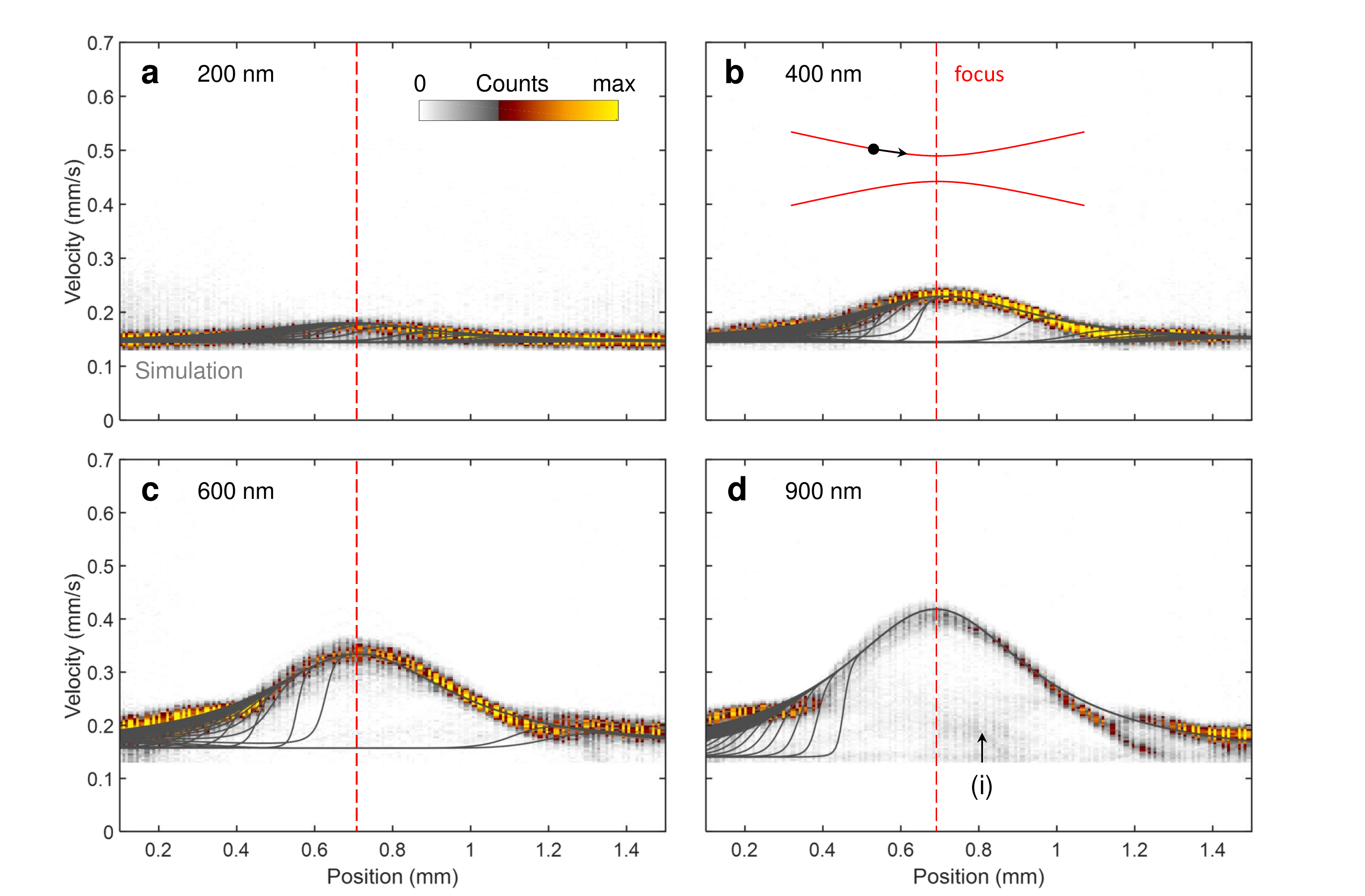}}
\caption{Histogram of measured velocities as a function of propagation distance and simulated particle trajectories (solid lines) for polystyrene particles with nominal diameters of (a) 200 nm, (b) 400 nm, (c) 600 nm, and (d) 900 nm.  With increasing diameter the optical forces and, in turn, the particle velocities in the focus region (dashed line) increase.  }
\end{figure}

\begin{equation}\label{eq:newton}
  \bm v(\bm r)=\bm v_{\rm fluid}+\frac{{\bm F}_{\rm opt}(\bm r)}{6\pi\eta R}\,.
\end{equation}
\noindent We emphasize that our model contains no free parameters, and all laser, fluid, and nanoparticle parameters can be inferred from experiment, as detailed in Appendix A.  

Figure~2c shows a few selected trajectories computed with our model.  Nanoparticles that are trapped move along the intensity maxima of the vortex beam, while nanoparticles that are not trapped are transported by the fluid without any significant deflection in the transverse direction.  From our simulations we can extract the velocities $v_z(z)$ as a function of propagation length $z$, see solid lines in Fig.~2d.  We can distinguish three types of trajectories.  First for particles that are trapped from the beginning (left) and are accelerated by the optical forces throughout, their $v_z(z)$ curves are in almost perfect agreement with the maxima of the experimental velocity histograms.  Second for particles that are initially away from the intensity maxima, are deflected by optical forces while propagating down the flow channel, and become finally trapped.  Third for particles that are initially even further away from the intensity maxima and are never trapped.  As only nanoparticles situated in high-field regions of the laser beam can efficiently scatter light, in our experiments we only observe trapped particles while non-trapped particles remain dark.

\section{Results}

Figure 4 shows velocity histograms for standard polystyrene particles with approximately monomodal size distributions and diameters of (a) 200 nm, (b) 400 nm, same as Fig.~2, (c) 600 nm, and (d) 900 nm.  One observes that all trapped particles become accelerated towards the focus region (dashed lines).  With exception of the 200 nm sample, where the velocity changes are relatively small for the chosen fluid velocity and laser intensity, we observe substantial velocity enhancements up to a factor of three.  The maximal velocities in the focus region increase with increasing diameters because of the stronger optical forces exerted on larger particles.  Comparison with the simulation results (solid lines) for trapped particles reveals very good agreement with experiment throughout.  For the larger particles we occassionally register spurious velocity counts, see region indicated with (i) and the bimodal velocity distribution at larger propagation distances.  Analyzing in more detail the corresponding particle trajectories, we can trace back such counts to artifacts in our detection algorithm, which are caused by overmodulation and defocusing of the camera.  Future work will address this point in more detail.

We finally comment on the capabilities of our OF2i scheme to measure particle size distributions.  Quite generally, as we can detect each individual particle passing through the focus region and can extract its velocity, we expect that we can also reliably infer a number based size distribution of the entire particle ensemble.  For small particles, say with diameters below 200 nm, the optical scattering forces in $z$ direction are too weak for a meaningful extraction of the particle size based on trajectory analysis, and we therefore instead have to resort to the scattering cross section for the size estimation.  This approach is similar to other, more  standard schemes and will not be further discussed in the following, although we have proven its applicability for a variety of unrelated samples.  To better understand the size extraction based on optical forces, in Fig.~5b we report simulation results for selected trajectories and for different nanoparticle sizes: with increasing particle diameter more particles become trapped in the transverse direction by the vortex laser beam, due to the increase of the optical forces with increasing particle volume, and are subsequently transported into the focus region and out of it.  Suppose that far away from the focus, where the optical forces are negligible, the particles start at position $(x,0,z_0)$ with velocity $(0,0,v_{\rm fluid})$.  We can now compute the particle trajectories from $z_0$ towards the center region $z_{\rm focus}$ using our velocity model of Eq.~\eqref{eq:newton}.  Figure~5a reports the velocities $v_{\rm focus}$ in the focus as a function of the impact parameter $x$ of the starting position.  Two important conclusions can be drawn from the figure.  First, the maximal velocities $v_{\rm focus}$ increase monotonously with the particle size, due to the related increase of optical forces.  For large particles Mie resonances might lead to a non-monotonic dependence, but we checked that this is not the case for the parameters and sphere diameters considered here.  Second, with the possible exception of the smallest particles, $v_{\rm focus}$ does not noticeably depend on the starting position~$x$.  This does not come unexpectedly.  Once the particles become trapped in the transverse direction, the gradient forces steer all particles in a completely similar fashion along the intensity maxima of the focused laser beam.  We can thus define a cutoff parameter $x_{\rm cut}(d)$ for particles with a diameter $d$ that become optically trapped.  The cutoff parameter is indicated on the left-hand side of Fig.~5b, particles that start with $x$ values larger than $x_{\rm cut}(d)$ are never trapped by the vortex beam and are transported down the fluid channel without scattering any light.  The

\begin{figure}[t]
\centerline{\includegraphics[width=\textwidth]{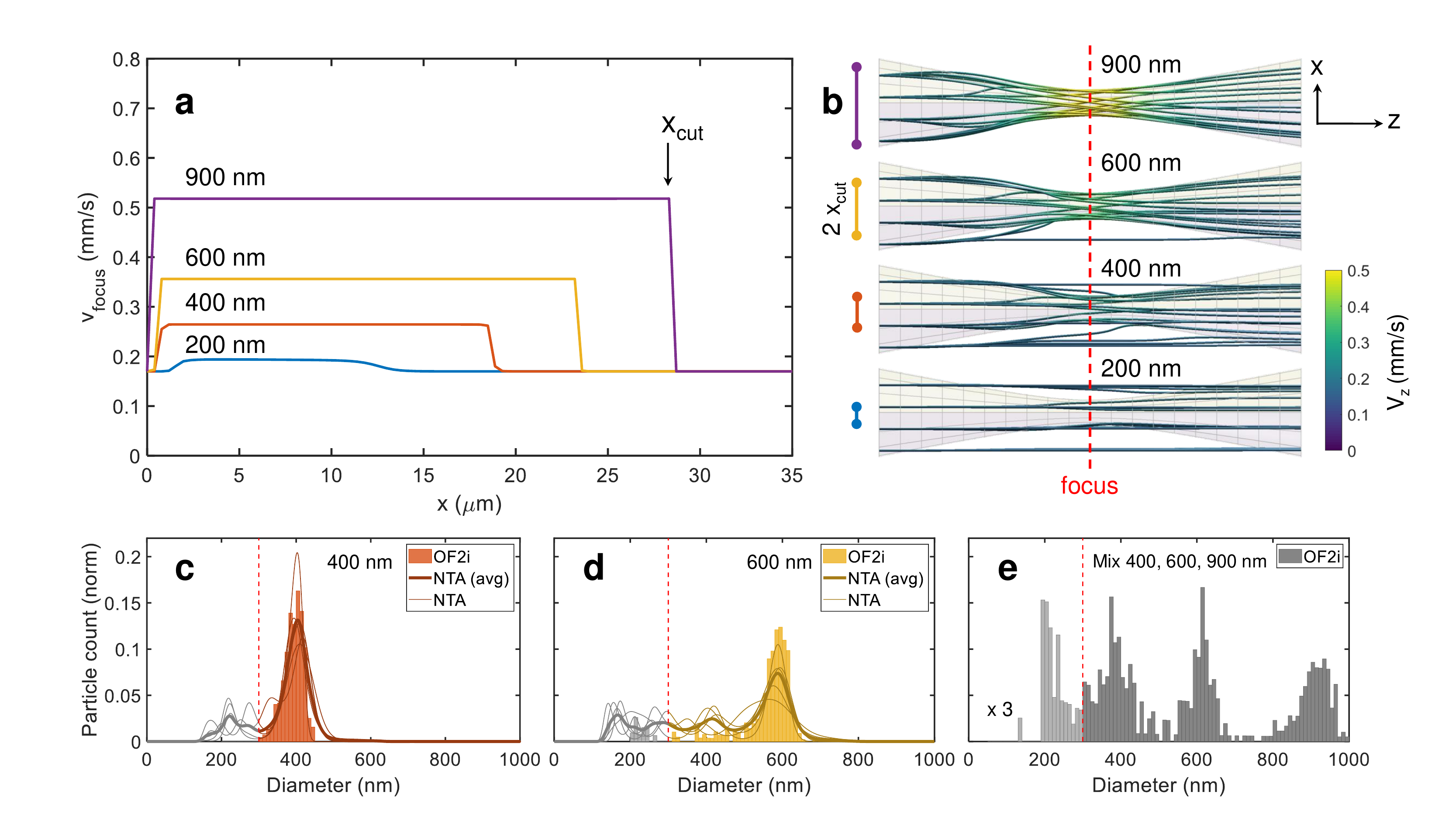}}
\caption{(a) Maximal velocity $v_{\rm focus}$ in the focus region as a function of the transverse direction~$x$ perpendicular to the beam axis $z$, for different particle diameters and computed within Mie theory.  With increasing diameter, the velocities $v_{\rm focus}$ increase monotonously as well as the cutoff parameter $x_{\rm cut}$, which is a measure for the number of nanoparticles trapped by the vortex beam.  (b) Selected trajectories for different nanoparticle diameters, the range of $x_{\rm cut}$ is indicated on the left.  (c,d) Particle size distribution for nanospheres with diameter of 400 nm and 600 nm, respectively.  The sum of the OF2i counts above 300 nm (see dashed line) is normalized to one and the bin width of the histograms is 10 nm.  The thin, solid lines show results of five independent NTA measurements, the thick lines reports the average.  (e) Particle size distribution for a mixture of polystyrene particles with diameters of 400 nm, 600 nm, and 900 nm.}
\end{figure}

These findings can now be combined with the fact that only particles in regions of considerably strong laser fields can scatter light, and can thus be detected by our microscope setup.  The active volume sampled by the OF2i scheme in a time interval $t_{\rm meas}$ becomes
\begin{equation}
  V_{\rm active}(d)=\Big[\pi x_{\rm cut}^2(d)\Big]v_{\rm fluid}t_{\rm meas}\,,
\end{equation}
where the term in brackets is the cross section in the transverse direction, and $v_{\rm fluid}t_{\rm meas}$ is the size of the sampling volume along the propagation direction.  The active volume corrects for the fact that larger particles are trapped more easily and are observed more frequently in comparison to smaller particles.  For $N_{\rm meas}$ velocity counts in a small diameter range, the particle density can be obtained through $n=\nicefrac{N_{\rm meas}}{V_{\rm active}}$.  

Figure~5(c,d) shows the particle size distributions for the standard particles with diameters of 400 nm and 600 nm, respectively, as measured within OF2i and computed according to the scheme described above.  To be on the safe side, we discard particles with a diameter of less than 300 nm (which can be analyzed using the scattering intensities) and normalize the distributions such that the sum of the histogram values gives one (we use a bin width of 10 nm).  One observes a relatively narrow size distribution that is peaked around the nominal values.  We compare our results with independent measurements performed with the nanoparticle tracking analysis (NTA) using the same standard particles, where the thin lines report individual results for approximately 70 particles each, and the solid lines give the average.  We found that the actual NTA distributions somewhat depend on the chosen parameters for the NTA software~\cite{filipe:10}, although in general the OF2i and NTA measurements are in good agreement.  It is known from the literature~\cite{bachurski:19} that NTA measurements often give larger nanoparticle spreads in comparison to transmission electron microscopy (TEM) experiments.  Our results suggest that a similar overestimation occurs in OF2i, but further work is needed to clarify this point in more detail.  Finally, in Fig.~5(e) we report results for a measurement with a mix of the previously analyzed spheres with diameters of 400 nm, 600 nm, and 900 nm, where the extracted size distribution of particles is in good agreement with the expected sum of distributions for the individual ensembles.  The counts below 300 nm are probably due to particles that are not fully trapped and artifacts of our velocity detection scheme, both points to be optimized in future software developments.

\begin{figure}[t]
\includegraphics[width=0.65\textwidth]{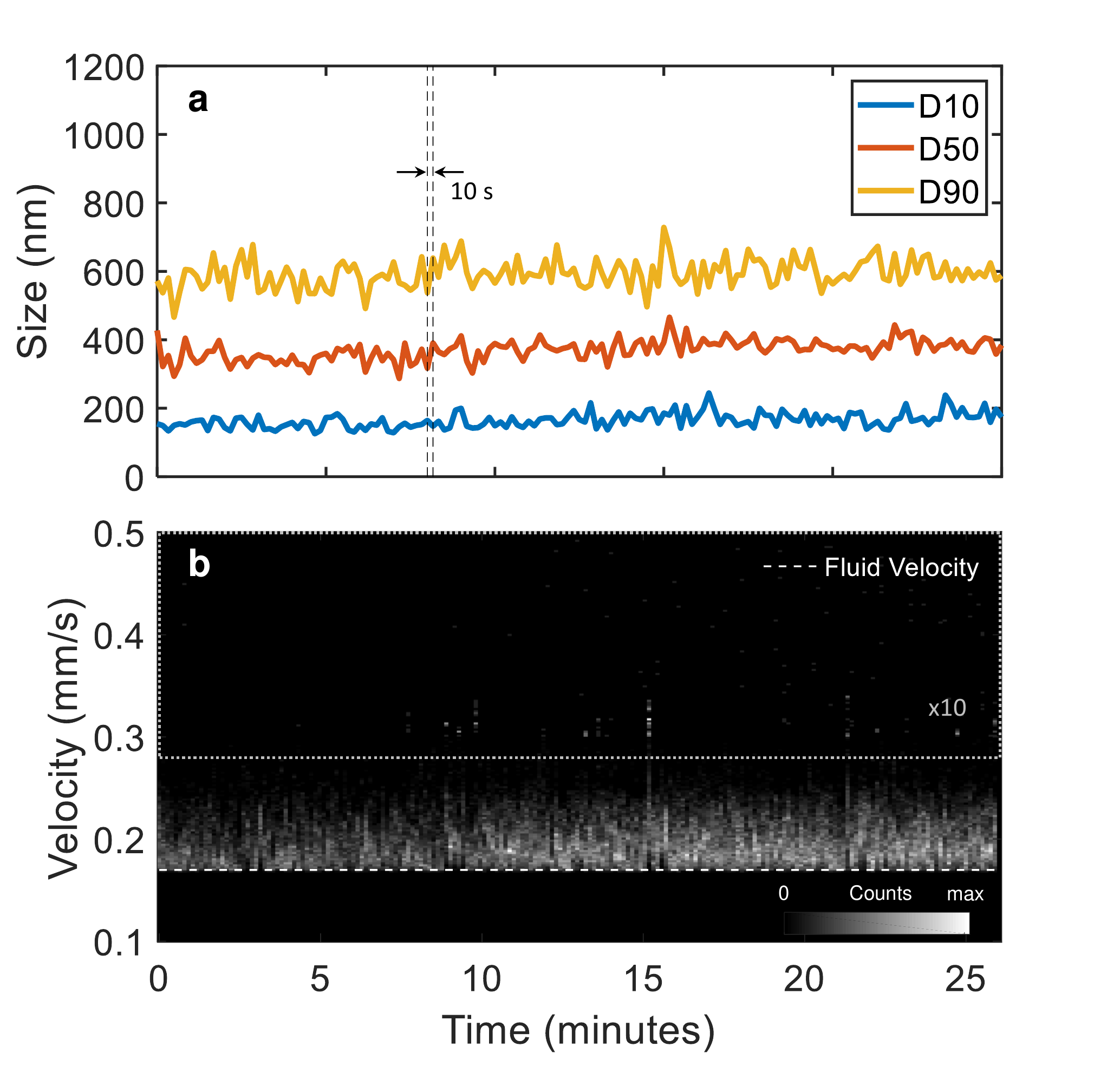}
\caption{Continuous monitoring of an oil-in-water emulsion process, as obtained in cooperation with one of our industry partners. (a) D-values as a function of time.  The values are extracted from the cumulative particle size distribution which is accumulated over a time period of ten seconds, and represent the diameter below which 10\%, 50\% and 90\% of all particles are detected.  With increasing particle size, the D-values increase according to the size distribution.  (b) Corresponding velocity histograms.  The dashed line indicates the flow velocity, the counts in the dotted rectangle have been multiplied by a factor of ten for better visibility of rare detections of large particles.}\label{fig:monitoring}
\end{figure}

We are currently implementing our OF2i measurement scheme in cooperation with one of our industry partners for a continuous monitoring of an established process for producing oil-in-water emulsions.  A highly concentrated sample is diluted to comply with our experimental setup, and is continuously pumped through a measurement cell at constant flow rate and laser power.  Figure~\ref{fig:monitoring}(a) reports time-averaged D-values, which are derived from the cumulative particle size distributions, each averaged over ten seconds.  D-values indicate the diameter below which 10\%, 50\% and 90\% of all particle diameters are measured, and serve as a process feedback parameter that directly reflects the emulsion quality.  Figure~\ref{fig:monitoring}(b) shows the histogram of velocities $v_{\rm{focus}}$ in the focus of the beam used for the calculation of the particle size distribution (spherical particles, refractive index of $n_p = 1.47$).  In the figure we enhance the counts within the dotted box by a factor of ten for better representation of low concentrated particles, where the objective of our control is to keep the concentration of large particles as low as possible.  Our results confirm that OF2i is capable of a real-time monitoring of particle size distributions in the context of process dynamics.

\section{Summary and Outlook}

In conclusion, we have presented a scheme for the real-time determination of particle size distributions with high throughput, which we term optofluidic force induction (OF2i).  It combines the measurement capabilities of light, which are at the heart of Brownian-motion based characterization schemes, with the manipulation capabilities of light through optical forces and torques.  Rather than trapping particles in all three dimensions, as routinely done in optical tweezers, we exploit optical forces for the trapping in the directions perpendicular to the propagation direction of a weakly focused laser only, and use the force in the parallel direction for a controlled nanoparticle motion through the focus region and the determination of particle sizes based on the velocity changes induced by the optical forces.  In our approach we currently extract an average nanoparticle diameter only, similar to Brownian motion based schemes, and additionally require an estimate for the refractive index of the particles.  This is usually not a major limitation, as for most applications the nanoparticle composition is approximately known and reliable estimates for the refractive index are at hand. For the future we plan to correlate the particle velocities with scattering data, potentially also in combination with spectroscopy, in order to obtain most detailed information about the nanoparticles.  We have already observed that light fields exert an optical torque on elongated particles, which are brought to a rotation that can be monitored.  This could allow us to additonally obtain more detailed information about the nanoparticle geometries.  With these potential improvements, OF2i will provide a flexible work bench for numerous pharmaceutical and technological applications, as well as for medical diagnostics.

\section*{Acknowledgements}

This work was supported in part by the Austrian Research Promotion Agency (FFG) through project LightMatters 870710, the European Commission (EC) through project NanoPAT (H2020-NMBP-TO-IND-2018-2020, Grant Agreement number: 862583). We are grateful to Michael Peinhopf, Alexander Leljak, and Michael Schnur for the excellent technical input within the BRAVE Analytics team.  We thank the whole nano-medicine workgroup at the Gottfried Schatz Research Center, especially Karin Kornm\"uller, Johann Krebs, Anna Schachner-Nedherer, and Ivan Vidakovic for their cooperation and most helpful discussions.

\begin{appendix}

\section{}

\begin{figure}[t]
\includegraphics[width=0.7\textwidth]{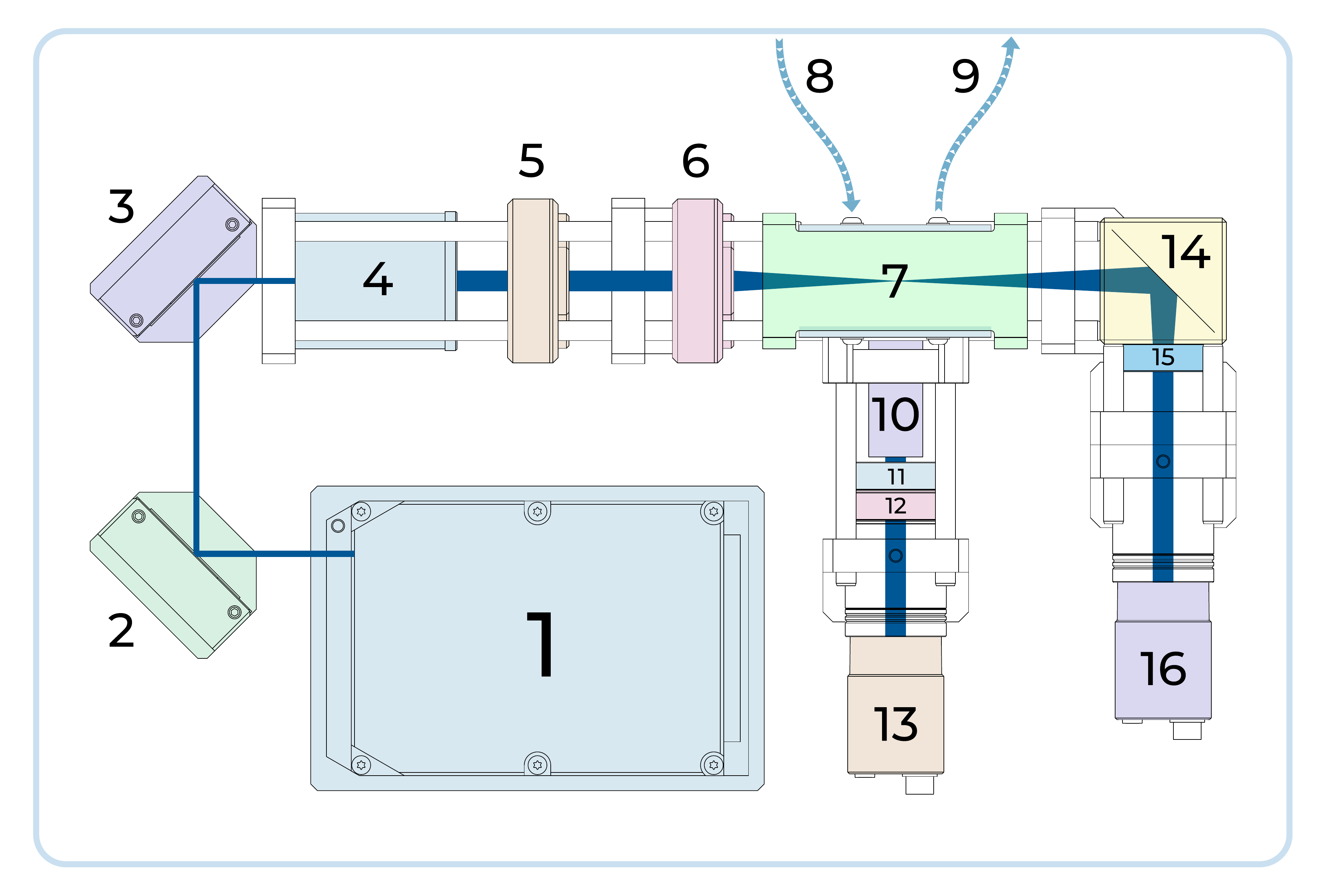}
\caption{Schematics for the OF2i setup. A Gaussian beam profile exits a laser source (1) at constant wavelength of 532 nm.  Passing mirrors (2,3), the beam is expanded by a beam expander (4) to fully illuminate a spiral phase plate (5).  A converging lens (6) is used to focus the vortex beam into a sample cell (7) where the light-matter interaction with the incoming sample (8) takes place.  Particles leaving the focal region are transported to waste via an outlet port (9).  The scattered light is magnified by an ultramicroscope setup (10-13) and recorded with a camera (13) at 200 fps.  In a second path (14-15), the illuminating beams cross-section is recorded using a second camera (16) to perform beam diagnostics and observe forward scattered light.}\label{fig:setup}
\end{figure}

In this Appendix we provide additional information about our experiments, the velocity extraction, and our simulations based on Mie theory.  A schematics of the OF2i setup is shown in Fig.~\ref{fig:setup}, see also the figure caption for the explanation of the various elements. In our experiments we use polystyrene samples (Thermo Scientific, 3000 Series NIST\textsuperscript{\texttrademark} traceable Nanosphere\textsuperscript{\texttrademark} Size Standards, $\approx\!1\%$ Polystyrene solids in 15 ml aqueous suspension) with nominal diameters of 200 nm ($\pm 4$ nm), 401 nm ($\pm 6$ nm), 600 nm ($\pm 9$ nm), and 903 nm ($\pm 12$ nm).  The Polystyrene Standards are diluted in ultrapure water type 1 (MilliQ\textsuperscript{\textregistered}), 0.02 $\mu$m micro-filtrated (inorganic sterile membrane filter Whatman Anotop\textsuperscript{\texttrademark}25, diameter: 25 mm, pore size: 0.02 $\mu$m), dilution factors for the pure samples are 1:20\,000 (200 nm), 1:50\,000 (400 nm), 1:20\,000 (600 nm) and 1:10\,000 (900 nm), and for the mix 1:100\,000 (200 nm), 1:20\,000 (400 nm), 1:20\,000 (600 nm) and 1:10\,000 (900 nm).   

For the velocity determination, we start with the waterfall diagrams, as exemplary shown in Fig.~2, and analyze for a given position the image within the range of $\pm$20 pixels.  In a second image we plot a straight line with gradient angle $\theta$ and with a Gaussian profile in the transverse direction ($\sigma=2$), and choose $\theta$ such that the overlap of the two images becomes maximal.  From $\theta$ we can compute the particle velocity at a specific position.  Particle positions are obtained from the raw video data in pixel format and transformed according to a developed ray-tracing model into a local coordinate system.  This conversion is done using the calculated constant of $1.18 \ \mu\textrm{m/pixel}$.

In our simulations, the fluid velocity $v_{\rm fluid}$ is computed from the maximum of the parabolic velocity profile of Hagen Poiseuilles's equation for a capillary with a diameter of 1 mm and for a dynamic viscosity of $\eta=9.544 \times 10^{-4}$~Pa\,s at $T = 22^{\circ}\mathrm{C}$.  We always take the maximum velocity of the flow profile, since particles deviate only slightly from the capillary center along the beams profile compared to the capillary diameter when they are optically trapped.  The flow velocity is calculated via the capillary's cross section together with the flow rate given in $\mu$L/s.  The refractive index of Polystyrene particles is $n_p=1.59$ and  $n_b = 1.33$ for the surrounding background medium (water) at a wavelength of $\lambda=532$ nm~\cite{sultanova:09}.  The effective focal length of the focused laser beam as well as the topological charge $m=2$ are chosen in accordance with experiment.
%The effective focal length of the focused laser beam is  $f = 50.8$ mm with a Gaussian beam waist of $\omega_0 = 1.8$ mm for the input beam, where the M$^2$ value of the laser light source is set to $M = 1$.  The Laguerre-Gaussian beam has a topological charge of $m_{\text{LG}} = 2$.  These parameters determine the Gaussian beam waist $\omega_1 = f\lambda / (\pi \omega_0)$ in the focal region.  
Within optical force calculations, we truncate the expansion of the electromagnetic fields at a maximum angular degree of $\ell_{max} = 20$ to ensure accurate results for the presented size range while minimizing the computational cost.  We checked that our results did not change noticeably when further increasing $\ell_{max}$.

\end{appendix}

%\bibliography{../of2i}

\end{document}